\documentclass[aps,amssymb,floatfix,prd,amsmath,preprintnumbers]{revtex4}
\usepackage{epstopdf}
\usepackage{capt-of}
\usepackage{graphicx}  
\usepackage{dcolumn}   
\usepackage{bm}
\begin{document}
\title{\bf Cosmological models with a hybrid scale factor in an extended gravity theory}

\author{B. Mishra \footnote{Department of Mathematics, Birla Institute of Technology and Science-Pilani, Hyderabad Campus, Hyderabad-500078, India, E-mail:bivudutta@yahoo.com },  S. K. Tripathy\footnote{Department of Physics, Indira Gandhi Institute of Technology, Sarang, Dhenkanal, Odisha-759146, India, E-mail:tripathy\_ sunil@rediffmail.com} and Sankarsan Tarai \footnote{Department of Mathematics, Birla Institute of Technology and Science-Pilani, Hyderabad Campus, Hyderabad-500078, India, E-mail:tsankarsan87@gmail.com},
}

\affiliation{ }

\begin{abstract}
A general formalism to investigate  Bianchi type $VI_h$ universes is developed  in an extended theory of gravity. A minimally coupled geometry and matter field is considered with a rescaled function of $f(R,T)$ substituted in place of the Ricci scalar $R$ in the geometrical action. Dynamical aspects of the models are discussed by using a hybrid scale factor that behaves as power law in an initial epoch and as an exponential form at late epoch. The power law behaviour and the exponential behaviour appear as two extreme cases of the present model. 
\end{abstract}

\maketitle

\section{Introduction:}

The prediction of late time cosmic acceleration happens to be an important discovery in the late part of the twentieth century. This discovery encouraged the experimentalist to come out with independent outcomes on the claim of accelerated expansion of the universe. On the classical cosmological front, the accelerated expansion occurs mostly due to a less known energy form called dark energy (DE) component. DE is considered as a fluid with negative pressure and accounts for about two-third (approximately  68 percent) of the total cosmic energy budget (Planck data). The DE can be described  with an equation of state (EoS) parameter $\omega=\frac{p}{\rho}$, where $p$ is pressure and $\rho$ is energy density. The cosmic expansion is predicted to be accelerating for $\omega < -\frac{1}{3}$. Cosmological constant in an $\Lambda$CDM model predicts the EoS to be $\omega = -1$. Moreover when $\omega<-1$, phantom dark energy dominates and may lead to a finite time future singularity called big rip singularity \cite{Cald03, Nojiri05a, Brevik13}. The cosmological constant is a prime candidate of DE and represents the quantum vacuum energy density.  The $\Lambda$CDM model is a successful model and can explain many observed phenomena. However this model is plagued by the fine tuning and coincidence problems. Also, DE becoming increasingly elusive as its origin and nature are yet to be understood. This has lead to the development of modified theory of gravity to explain the late time cosmic acceleration. Of late, modified theories of gravity are taking the place of General Relativity(GR) to explain the late time cosmic dynamics without considering any dark energy components in the matter field. \\

In the last few decades, many generalization of GR have been proposed. A lot of research has been done to address instability problems with the help of GR; however to explain the cosmological and astrophysical phenomena in a satisfactory way in the presence of dark matter, the instability problem is no longer helpful. Particularly the observational evidence of expanding universe has put theoretical cosmology into crisis. Since there is limitations of General Relativity (GR) on large scale, astrophysicists turn their attention to modified theories. With the modifications of the Einstein-Hilbert action, different modified theories have been proposed. One may refer to some good reviews on these extended theories of gravity \cite{Cappo2011, Nojiri2007, Nojiri2011, Bamba2012, Clifton2012, Soti2010}.  Many authors have investigated different issues related to cosmology and astrophysics in modified gravity theories such as $f(R)$ gravity, $f(T)$ gravity and $f(G)$ gravity \cite{Poplawski06, Sharif15, Cappo2007, Myrzakulov11, Dent11, Harko14, Li07, Kofinas14, Nojiri04, Nojiri05} . Bianchi type models with anisotropic spatial sections are interesting in the sense that they are more general  than the Friedman models. Even though there is a strong debate going on the viability of Bianchi type models \cite{Sadeh16}, these models can be useful in the description of early inflationary phase and with suitable mechanism can be reduced to isotropic behaviour at late times. In the framework of $f(R)$ gravity, Shamir and co authors have investigated different aspects of Bianchi type models \cite{Sharif2009, Sharif2010, Shamir2010}. Recently Tripathy and Mishra have obtained anisotropic cosmological solutions in $f(R)$ gravity \cite{Tripathy16}. Momeni and Gholizade obtained some cylindrically symmetric solutions in this theory \cite{Momeni2009}.


With a further modification to GR, Harko et al.\cite{Harko11} proposed  the $f(R,T)$ theory basing on a coupling between matter and geometry.  In this modified theory, the four-dimensional Einstein-Hilbert action is written as

\begin{equation} \label{eq:1}
S=\frac{1}{16\pi}\int d^4x\sqrt{-g} f(R,T)+\int d^4x\sqrt{-g} \mathcal{L}_m,
\end{equation}
where $f(R,T)$ in the action is a function of $T(=g_{ij}T^{ij})$ and Ricci scalar $R$. $T^{ij}$ is the energy-momentum tensor. The matter Lagrangian $\mathcal{L}_m$ can be taken either as $\mathcal{L}_m=-p$ or as $\mathcal{L}_m=\rho$. Here, we will chose the first option for brevity. In this theory, Samanta \cite{Samanta13} has obtained exact solution of $f(R,T)$ gravitational field equations in Kantowski-Sachs space time. Moraes \cite{Moraes15} obtained some exact solutions in higher dimensional space-time. Zubair and Hassan \cite{Zubair16} have reconstructed cosmological models for Bianchi type I,III and Kantowski-Sachs solutions.  Shamir \cite{Shamir15} has obtained the exact solutions of the field equation in LRS Bianchi type I space time under some adhoc conditions whereas Shabani and Ziaie \cite{Shabani17} have investigated the stability of the Einstein static universe in $f(R,T)$ gravity. Mishra et al. \cite{Mishra15} have constructed non-static cosmological models in both the linear and quadratic form of the trace. In another work, Mishra and Vadrevu \cite{Mishra17} constructed some cosmological models in the linear and quadratic form of Ricci scalar $R$ in $f(R,T)$ gravity. Several investigations were made in this theory to study the dynamical aspects of the anisotropic cosmological models \cite{Mishra16, Agrawal17,  Mishra17a,Mishra17b,Mishra17c, Aktas17, Ilyas2017}.\\

The objective of the paper is to study Bianchi type $VI_h$ ($BVI_h$) models within the framework of an extended theory of gravity as proposed by Harko et al. \cite{Harko11}. Our motivation in the present work is to develop a general formalism to investigate such anisotropic models with a time varying deceleration parameter simulated by a hybrid scale factor (HSF). The concept of HSF has already been conceived in our earlier works \cite{BM15, SKT14, BM18}.  Time varying deceleration parameter is required to explain a transition of the universe from a decelerated phase to an accelerated phase at a recent epoch. The paper is organised as follows: In section II, we have developed the basic field equations in the framework of $f(R,T)$ gravity and derived the relevant geometrical parameters. In section III, the dynamics of the model is presented along with the energy conditions. The summary and conclusion are given in section IV.

\section{Basic Field Equations within $f(R,T)$ framework}
In this section, we discuss briefly the formalism developed to investigate certain models in a minimally coupling $f(R,T)$ theory. We consider a Bianchi $VI_h$ ($BVI_h$) space time
\begin{equation}\label{eq:2}
ds^{2}=dt^{2}-A^{2}dx^{2}-B^{2}e^{2x}dy^{2}-C^{2}e^{2hx}dz^{2},
\end{equation}
where $A,$ $B$ and $C$ are functions of cosmic time $t$. The exponent $h$ in the space-time is self-governing constant and can take integral values as $-1,0,1$. In this work, we have considered the self-governing constant $h=-1$ because of the importance of the metric that envisages an isolated universe with null total energy and momentum \cite{Tripathy15,SKT16}. The matter field is considered through an energy momentum tensor    
\begin{equation}  \label{eq:3}
T_{ij}=(p+\rho)u_iu_j - pg_{ij}-\rho_B x_i x_j.
\end{equation}
Here $u^{i}x_{i}=0$ and $x^{i}x_{i}=-u^{i}u_{i}=-1$. In a co moving coordinate system, $u^{i}=\delta_0^i$ is the four velocity vector of the fluid. $x^{i}$ represents the direction of anisotropic fluid (here $x$-direction) and is orthogonal to $u^{i}$. The energy density $\rho$ is composed of energy density due to the perfect fluid and that due to an anisotropic fluid $\rho_B$. For the modified gravity model, we consider here a minimal coupling of geometry and curvature assuming $f(R,T)=f(R)+f(T)$. Following our earlier works, we can write the field equations as \cite{Mishra16,Mishra17b}, 

\begin{equation} \label{eq:4}
f_R (R) R_{ij}-\frac{1}{2}f(R)g_{ij}+\left(g_{ij}\Box-\nabla_i \nabla_j\right)f_R(R)=\left[8\pi+f_T(T)\right]T_{ij}+\left[pf_T(T)+\frac{1}{2}f(T)\right]g_{ij}
\end{equation}
where $f_R=\frac{\partial f(R)}{\partial R}$ and $f_T=\frac{\partial f(T)}{\partial T}$ are the respective partial differentiations. 

A specific choice of $f(R,T)=\lambda(R+T)$ leads to the field equations \cite{Mishra17b}

\begin{equation}\label{eq:5}
G_{ij}=\left(\frac{8\pi+\lambda}{\lambda}\right)T_{ij}+\Lambda(T)g_{ij}, 
\end{equation}
where $\lambda$ is a non zero scaling factor that rescales the usual field equations in GR. $G_{ij}=R_{ij}-\frac{1}{2}Rg_{ij}$ is the Einstein tensor. The factor $\Lambda(T)=p+\frac{1}{2}T$ appearing in the field equation \eqref{eq:5} may be identified with a time dependent effective cosmological constant. Here $\Lambda(T)$ depends on the matter content and helps in providing an acceleration. Even though, the field equations in \eqref{eq:5} have the same mathematical form of GR with a time varying constant, it cannot be reduced to GR because of the non vanishing quantity $\lambda$. However, eq.\eqref{eq:5} is a rescaled generalisation of GR equations.
The field equations \eqref{eq:5} of the modified $f(R, T)$ gravity theory, for Bianchi type $VI_{h}$ space-time can be explicitly written as 

\begin{eqnarray}
\frac{\ddot{B}}{B}+\frac{\ddot{C}}{C}+\frac{\dot{B}\dot{C}}{BC}+\frac{1}{A^{2}} &=& -\alpha\left(p-\rho_{B}\right)+\frac{\rho}{2},\label{eq:6}\\
\frac{\ddot{A}}{A}+\frac{\ddot{C}}{C}+\frac{\dot{A}\dot{C}}{AC}-\frac{1}{A^{2}} &=& -\alpha p+\left(\frac{\rho_{B}+\rho}{2}\right),\label{eq:7}\\
\frac{\ddot{A}}{A}+\frac{\ddot{B}}{B}+\frac{\dot{A}\dot{B}}{AB}-\frac{1}{A^{2}} &=& -\alpha p+\left(\frac{\rho_{B}+\rho}{2}\right),\label{eq:8}\\
\frac{\dot{A}\dot{B}}{AB}+\frac{\dot{B} \dot{C}}{BC}+\frac{\dot{A}\dot{C}}{AC}-\frac{1}{A^{2}} &=& \alpha \rho-\left(\frac{p-\rho_{B}}{2}\right),\label{eq:9}\\
\frac{\dot{B}}{B} &=&\frac{\dot{C}}{C},\label{eq:10}
\end{eqnarray}
where $\alpha=\frac{16\pi+3\lambda}{2\lambda}$ and ordinary time derivatives are denoted by overhead dots. Directional Hubble parameters for the anisotropic model are $H_x=\frac{\dot{A}}{A}$, $H_{y}=\frac{\dot{B}}{B}$ and $H_z=\frac{\dot{C}}{C}$. In view of eq. \eqref{eq:10}, we have $H_y=H_z$. Assuming $H_x = kH_z$, for $k\neq1$, one may obtain a mean Hubble parameter as $H=\frac{\dot{\mathcal{R}}}{\mathcal{R}}=\left(\frac{k+2}{3}\right)H_{z}$, where $\mathcal{R}$ is the radius scale factor. 

The  field equations  \eqref{eq:6}-\eqref{eq:10} can be expressed in terms of $H$ as   

\begin{eqnarray}
\frac{6}{(k+2)} \dot{H}+ \frac{27}{(k+2)^{2}} H^{2}+\frac{1}{A^{2}} &=& -\alpha(p-\rho_{B})+\frac{\rho}{2},\label{eq:11}\\
\frac{3(k+1)}{(k+2)} \dot{H}+ \frac{9(k^{2}+k+1)}{(k+2)^{2}} H^{2}-\frac{1}{A^{2}} &=& -\alpha p +\left(\frac{\rho+\rho_{B}}{2}\right),\label{eq:12}\\
\frac{9(2k+1)}{(k+2)^{2}} H^{2}-\frac{1}{A^{2}} &=& \alpha \rho-\left(\frac{p-\rho_{B}}{2}\right).\label{eq:13}
\end{eqnarray}

Two different approaches can be adopted to get some viable cosmological models from the above field equations. Firstly, one can chose to assume a physically acceptable equation of state and then the dynamics of the universe is studied. On the other hand, basing upon the observationally chalked out dynamics of the universe, one may consider a presumed expansion behaviour and then the background cosmology is investigated. In the present formalism, we consider a presumed dynamical behaviour of the universe according to recent observational data and investigate the evolutionary behaviour.  

Some relevant geometrical parameters such as the scalar expansion $\theta$, shear scalar $\sigma^2$, average anisotropy parameter $\mathcal{A}$ are expressed as
\begin{eqnarray}
\theta &=& (k+2)H_z,\label{eq:14}\\
\sigma^2 &=& \frac{1}{3}(k^2-2k+1)H_z^2,\label{eq:15}\\
\mathcal{A} &=& \frac{2}{3}\left(\frac{k-1}{k+2}\right)^2.\label{eq:16}
\end{eqnarray}

Of these geometrical parameters, $\theta$ and $\sigma^2$ depend on the cosmic dynamics where as the average anisotropic parameter depends only on $k$.

Other dynamical parameters that depend on the higher derivatives of the scale factors are the deceleration parameter (DP) $q=-1-\frac{\dot{H}}{H^2}$ and the jerk parameter $j=\frac{\dddot{\mathcal{R}}}{\mathcal{R}H^3}$. For the present model they are given by
\begin{eqnarray}
q &=& -1-\left(\frac{3}{k+2}\right)\frac{\dot{H_z}}{H_z^2},\label{eq:17}\\
j &=& \left(\frac{3}{k+2}\right)^2\frac{\ddot{H_z}}{H_z^3}-(2+3q).\label{eq:18}
\end{eqnarray}
One should note that, once the dynamical behaviour of $H_z$ is known, then the evolutionary aspects of these two parameters can be well assessed.
\section{Dynamics of the model}
From the field equations \eqref{eq:11}-\eqref{eq:13}, we will be able to obtain the expressions of the physical parameters of the model as

\begin{eqnarray}
p &=& \frac{6}{(1-4\alpha^{2})} \left[\frac{(k-1)+2\alpha(k+1)}{(k+2)}\frac{\ddot{\mathcal{R}}}{\mathcal{R}}+
\frac{(2k^2-4k-7)+2\alpha(2k^2+1)}{(k+2)^2}\left(\frac{\dot{\mathcal{R}}}{\mathcal{R}}\right)^2\right]-\frac{2\mathcal{R}^{-\frac{6k}{k+2}}}{1-2\alpha},\label{eq:19}\\
\rho &=& \frac{6}{(1-4\alpha^2)}\left[\frac{2}{k+2}\frac{\ddot{\mathcal{R}}}{\mathcal{R}}+ \frac{(5-2k)-6\alpha(2k+1)}{(k+2)^2}\left(\frac{\dot{\mathcal{R}}}{\mathcal{R}}\right)^2\right]+\frac{2\mathcal{R}^{-\frac{6k}{k+2}}}{1-2\alpha},\label{eq:20}\\
\rho_B &=& \frac{6}{(1-2\alpha)}\left[\left(\frac{k-1}{k+2}\right)\left(\frac{\ddot{\mathcal{R}}}{\mathcal{R}}+2\frac{\dot{\mathcal{R}}^2}{\mathcal{R}^2}\right)\right]-\frac{4\mathcal{R}^{-\frac{6k}{k+2}}}{1-2\alpha}.\label{eq:21}
\end{eqnarray}

Other dynamical features of the model are the EoS parameter $\omega$ and the effective cosmological constant $\Lambda$. Using the scale factors, these parameters are obtained as

\begin{eqnarray}
\omega &=&-1+(1+2\alpha)\left[\frac{3(k^2+3k+2)\frac{\ddot{\mathcal{R}}}{\mathcal{R}}+6(k^2-3k-1)\frac{\dot{\mathcal{R}}^2}{\mathcal{R}^2}}{6(k+2)\frac{\ddot{\mathcal{R}}}{\mathcal{R}}-3(2k-5)\frac{\dot{\mathcal{R}}^2}{\mathcal{R}^2}+(k+2)^2 \mathcal{R}^{-\frac{6k}{k+2}}-2\alpha \left[ 9(2k+1)\frac{\dot{\mathcal{R}}^2}{\mathcal{R}^2}-(k+2)^2 \mathcal{R}^{-\frac{6k}{k+2}}\right]}\right], \label{eq:22} \\
\Lambda &=& \frac{6}{(1+2\alpha)(k+2)}\left[\frac{\ddot{\mathcal{R}}}{\mathcal{R}}+2\frac{\dot{\mathcal{R}}^2}{\mathcal{R}^2}\right].\label{eq:23}
\end{eqnarray}

In the above equations, all the physical parameters are expressed in terms of the scale factor. Therefore, if the expansion history can be tracked by assuming a scale factor, then the background cosmology can be easily investigated.  It is almost conclusive from different observational data that, the cosmic acceleration is a recent phenomena and there must have occurred a transition from  deceleration to an accelerated one in recent past. One should note that a constant deceleration parameter can not explain such a characteristics of the expansion. In view of this a time varying DP (TVDP) is required to understand the present universe that can behave according to the early deceleration and late time acceleration . In other words, the deceleration parameter should be positive at some initial epoch and after a signature flipping at some point of time, it becomes negative to describe an accelerated universe. Such a deceleration parameter can be obtained by a hybrid scale factor (HSF), $\mathcal{R}=e^{at}t^b$, proposed in some earlier works \cite{BM15, SKT14, BM18}.  The time varying deceleration parameter as obtained from a HSF is given by  $q=-1+\frac{b}{(at+b)^{2}}$. It behaves as $q\simeq -1+\frac{1}{b}$ when $t\rightarrow 0$ and as $t\rightarrow \infty$, it becomes $q\simeq -1$. The positive constant parameters of HSF, $a$ and $b$ can be constrained from the cosmic transit behaviour. In a recent work, we have constrained $b$ from some physical and plausible arguments to be in a range $0\leq b\leq \frac{1}{3}$ \cite{BM15}. Also in a previous work, we have constrained $a$ in the range $0.075\leq a\leq 0.1$ according to the recent observational constraints  on transition redshift $0.4\leq z_t \leq 0.8 $ \cite{BM18}. However, in the present work, we chose $a=0.695$ and $b=0.085$ so that it can predict a transition redshift of $z_t=0.806$. This value of transition redshift has been obtained in a recent work \cite{Jesus17}. Similar results have also been obtained from an analysis of Hubble parameter data \cite{Farooq17}. In the Fig.1, we have plotted the deceleration parameter obtained from the HSF that shows a transition at a redshift $z_t=0.806$.

\begin{figure}[t]
\minipage{0.60\textwidth}
\includegraphics[width=85mm]{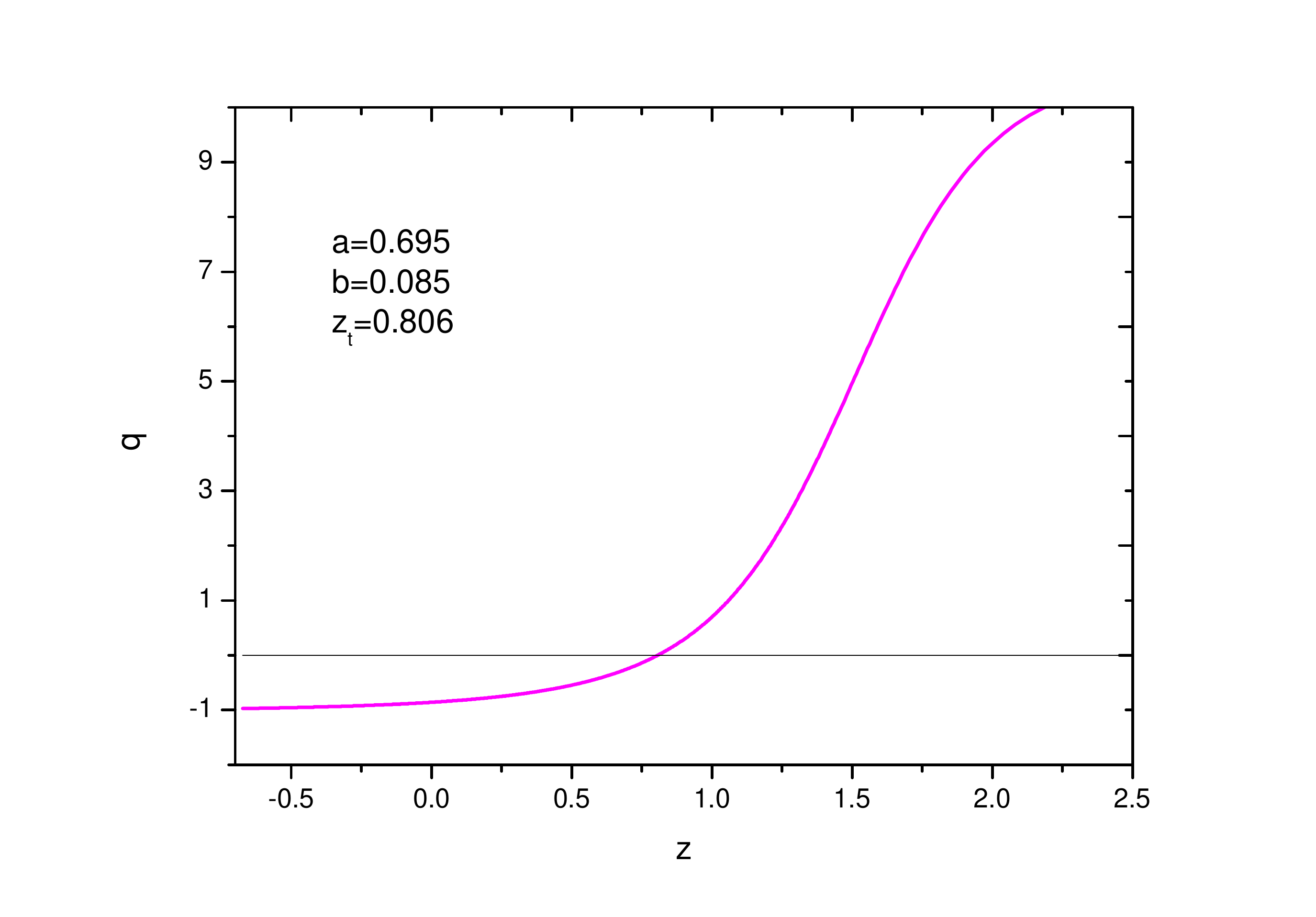}
\caption{Deceleration parameter $q$ for HSF.}
\endminipage
\end{figure}

Assuming the HSF, we can express the directional Hubble parameters as, $H_x=\frac{3k}{k+2}(a+\frac{b}{t})$  and $H_{y}=H_{z}=\frac{3}{k+2}(a+\frac{b}{t})$. The mean Hubble rate becomes $H=a+\frac{b}{t}$. The directional scale factors can be obtained as $A=e^{\frac{akt}{k+2}} t^{\frac{bk}{k+2}}$ and $B=C=e^{\frac{at}{k+2}}t^{\frac{b}{k+2}}$.  The kinematical parameters of the model with the presumed HSF  are  obtained as

\begin{eqnarray}
\Theta &=& 3a+\frac{3b}{t},\label{eq:24}\\
\sigma^{2} &=& 3\left(\frac{k-1}{k+2}\right)^2\left(a+\frac{b}{t}\right)^2.\label{eq:25}
\end{eqnarray}

It is now straight forward to obtain the expressions of the pressure $p$, energy density $\rho$ and the energy density of the anisotropic fluid $\rho_B$ from eqs. \eqref{eq:19}-\eqref{eq:21} using the HSF:

\begin{equation}\label{eq:26}
p=-\frac{6}{1-4\alpha^{2}}\left[\frac{b(3b\phi_2-\phi_1)+3at\phi_2(at+2b)}{(k+2)^2}\right]\frac{1}{t^2}-\frac{2}{1-2\alpha} e^{-\frac{6akt}{k+2}}t^{-\frac{6bk}{k+2}},
\end{equation}
\begin{equation}\label{eq:27}
\rho=\frac{6}{1-4\alpha^{2}}\left[ \frac{3b^2\phi_3-2b(k+2)+3at\phi_3(at+2b)}{(k+2)^{2}}\right]\frac{1}{t^2}+\frac{2}{1-2\alpha} e^{-\frac{6akt}{k+2}}t^{-\frac{6bk}{k+2}}, 
\end{equation}
\begin{equation}\label{eq:28}
\rho_{B}=\frac{6(k-1)}{1-2\alpha}\left[\frac{b(3b-1)+3at(at+2b)}{k+2}\right]\frac{1}{t^2}-\frac{4}{1-2\alpha} e^{-\frac{6akt}{k+2}}t^{-\frac{6bk}{k+2}}.
\end{equation}
In the above equations, we have redefined the constants as $\phi_1=(2-k-k^2)-2\alpha(k^2+3k+2)$, $\phi_2=(3+k-k^2)-2\alpha(k^2+k+1)$ and $\phi_3=3-2\alpha(2k+1)$.
Consequently, the EoS parameter $\omega$ and $\Lambda$ are obtained as
\begin{equation}\label{eq:29}
\omega=-1+3(1+2\alpha)\left[ \frac{3b[3b(k^2-k)-(k^2+3k+2)]+9(k^2-k)at(at+2b)}{3[3b^2\phi_3-2b(k+2)+3at\phi_3(at+2b)]+(1+2\alpha)(k+2)^2e^{-\frac{6akt}{k+2}}t^{\frac{2(k+2-3bk)}{k+2}}}\right] 
\end{equation}
\begin{equation}\label{eq:30}
\Lambda=\frac{6}{(k+2)(1+2\alpha)}\left[b(3b-1)+3at(at+2b)\right]\frac{1}{t^2}.
\end{equation}

\begin{figure}[t]
\minipage{0.60\textwidth}
\includegraphics[width=85mm]{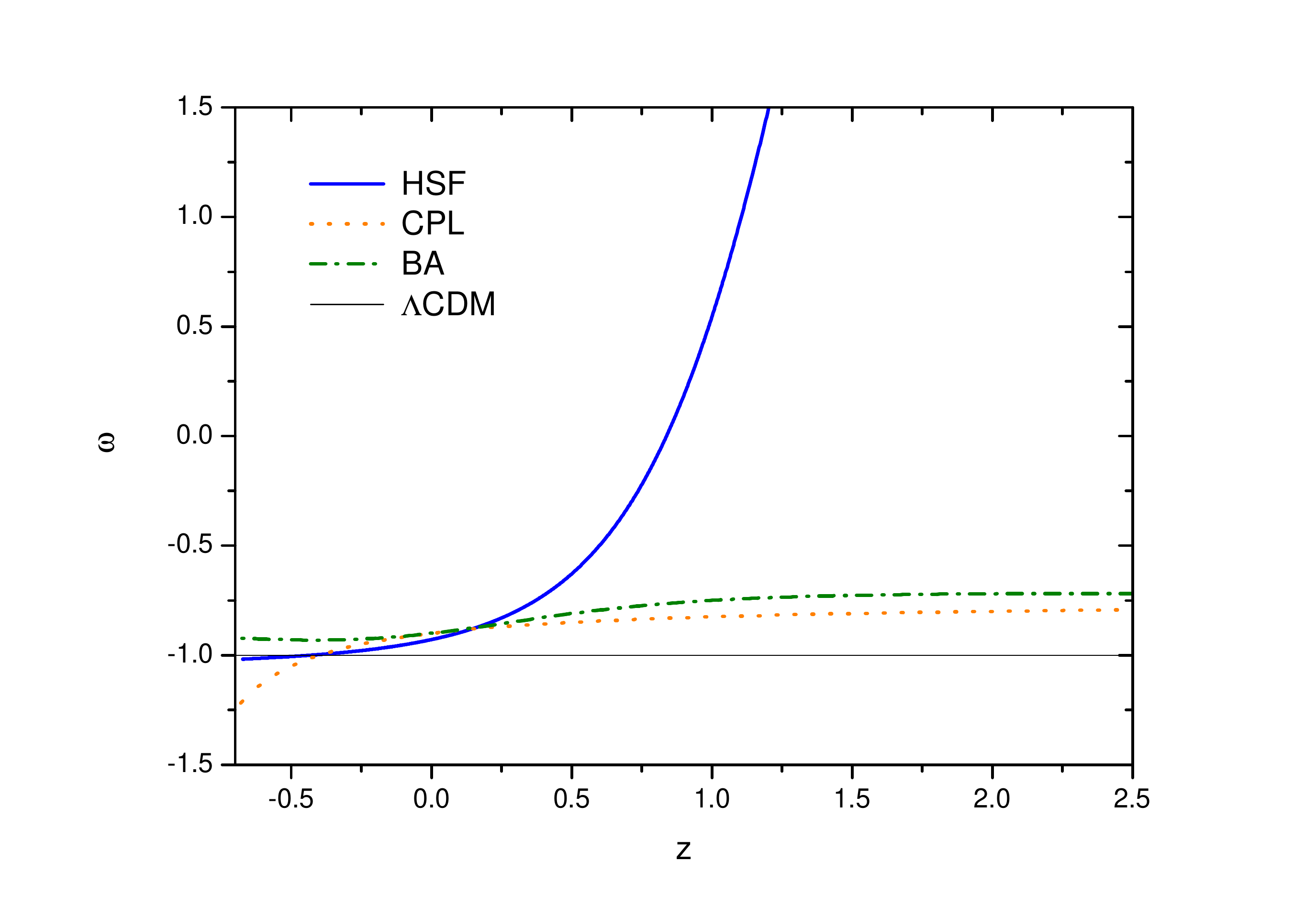}
\caption{Evolutionary behaviour of $\omega$ of HSF. The EoS parameters as calculated from the CPL parametrization (dotted curve), BA parametrization (dash-dotted curve) and the prediction of $\Lambda$CDM model ( solid black line) are shown for comparison.}
\endminipage
\end{figure}

In order to assess the dynamical aspects of the model, we have plotted the EoS parameter $\omega$ as a function of redshift in Fig.2. In the figures, we have fixed the anisotropic parameter as $k=1.1$. The HSF parameters as constrained from recent transition redshift data are considered for plotting the figure.  $\omega$ decreases from a positive value in the early phase to behave as a pure cosmological constant at late phase of cosmic time. The behaviour of the EoS parameter of HSF model has been compared with that of some well known EoS parametrizations such as the Chevallier-Polarski-Linder (CPL) \cite{CPL2001, CPL2003} and Barboza-Alcaniz(BA) \cite{BA12} parametrizations given as

\begin{eqnarray}
\textbf{CPL}~:~~ \omega(z) &=& \omega_0+\omega_a\frac{z}{1+z}\\
\textbf{BA}:~~ \omega(z) &=& \omega_0+\omega_a\frac{z(1+z)}{1+z^2},
\end{eqnarray}
where $\omega_0$ and $\omega_a$ are constants. The redshift $z$ is defined through $1+z=\frac{\mathcal{R}_0}{\mathcal{R}}$ where $\mathcal{R}_0$ is the scale factor at the present epoch.

In the low redshift region, predictions from these parametrizations are more or less the same as that of our model with HSF. However, at high redshift, EoS parameter from HSF rises with greater slope than these parametrizations. Since the HSF has a power law behaviour at early times and an exponential behaviour at late times, the same has been reflected in the figure. At late epoch, $\omega$ coincides with that of the $\Lambda$CDM model. At the present epoch, this model gives an $\omega=-0.929$ which is close to that of the $\Lambda$CDM model i.e  $\omega=-1$ which is consistent with the observational bounds. It is to mention here that, the EoS from HSF, may have different trajectories if we constraint the HSF parameters from some other physical basis but at late times all those trajectories overlap with that of $\Lambda$CDM model.

The interesting feature of the dynamical properties of the model with HSF is that, the expressions as obtained above are more general than the power law or the exponential expansion. It is worth to mention here that, in most of the cosmological models, authors use either a power law  scale factor or  an exponential one. These two behaviours appear as the two extreme cases of HSF. In fact, the dynamical evolution track of the EoS parameter for HSF lies in between the two extreme tracks predicted by a power law and an exponential scale factor. From the set of the equations \eqref{eq:26}-\eqref{eq:30}, we can always recover the relevant equations for the two extreme cases.

\subsection{Case-I: $a=0$}
The power law behaviour can be recovered from HSF, if we consider $a=0$ in the equations \eqref{eq:26}-\eqref{eq:30}, so that we can obtain the required expressions for pressure, energy density, energy density of the anisotropic fluid, equation of state and the effective cosmological constant:
\begin{eqnarray}
p &=& -\left[\frac{6b(3b\phi_2-\phi_1)}{(1-4\alpha^{2})(k+2)^2}\right]\frac{1}{t^2}-\frac{2}{1-2\alpha} t^{-\frac{6bk}{k+2}},\label{eq:31}\\
\rho &= &\left[ \frac{18b^2\phi_3-12b(k+2)}{(1-4\alpha^{2})(k+2)^{2}}\right]\frac{1}{t^2}+\frac{2}{1-2\alpha} t^{-\frac{6bk}{k+2}},\label{eq:32}\\ 
\rho_{B} &=& \left[\frac{6b(3b-1)(k-1)}{(1-2\alpha)(k+2)}\right]\frac{1}{t^2}-\frac{4}{1-2\alpha} t^{-\frac{6bk}{k+2}},\label{eq:33}\\
\omega &=& -1+9(1+2\alpha)\left[ \frac{b[3b(k^2-k)-(k^2+3k+2)]}{3[3b^2\phi_3-2b(k+2)]+(1+2\alpha)(k+2)^2t^{\frac{2(k+2-3bk)}{k+2}}}\right],\label{eq:34}\\
\Lambda &=& \left[\frac{6b(3b-1))}{(k+2)(1+2\alpha)}\right]\frac{1}{t^2}.\label{eq:35}
\end{eqnarray}

A substitution of $b=\frac{m}{3}$ in the above equations recovers the results obtained in an earlier work with the assumption of a power law scale factor behaving like $\mathcal{R}=t^\frac{m}{3}$\cite{Mishra17b}. The dynamical behaviour of this model will be the same as have been obtained in Ref. \cite{Mishra17b}.

\subsection{Case-II: b=0}
 
 If one considers $b=0$ in the HSF, de Sitter model with an exponential expansion can be achieved. In such a case the dynamical parameters of the model become
 
\begin{equation}\label{eq:36}
p=-\frac{18a^2\phi_2}{(1-4\alpha^{2})(k+2)^2}-\frac{2}{1-2\alpha} e^{-\frac{6akt}{k+2}},
\end{equation}
\begin{equation}\label{eq:37}
\rho=\frac{18a^2\phi_3}{(1-4\alpha^{2})(k+2)^2}+\frac{2}{1-2\alpha} e^{-\frac{6akt}{k+2}},
\end{equation}
\begin{equation}\label{eq:38}
\rho_{B}=\frac{18a^2(k-1)}{(1-2\alpha)(k+2)}-\frac{4}{1-2\alpha} e^{-\frac{6akt}{k+2}}.
\end{equation}
\begin{equation}\label{eq:39}
\omega=-1+\left[ \frac{27(1+2\alpha)(k^2-k)a^2}{9a^2\phi_3+(1+2\alpha)(k+2)^2e^{-\frac{6akt}{k+2}}}\right] 
\end{equation}
\begin{equation}\label{eq:40}
\Lambda=\frac{18a^2}{(k+2)(1+2\alpha)}.
\end{equation}
 
In the exponential model, $\omega$ increases from some higher negative value in the phantom domain in an initial epoch to behave like a cosmological constant at a late epoch. The effective cosmological constant becomes a time independent quantity for this model. 

Different energy conditions like the Null energy condition (NEC) ($\rho+p \geq 0$), Strong energy condition (SEC) ($\rho+3p$) and the Dominant energy condition (DEC) ($\rho-p$) can also be investigated in the model with HSF. Different energy conditions as obtained in this model are

\begin{eqnarray}
\textbf{NEC} &:& \rho+p =\frac{6}{1-4\alpha^2}\left[\frac{3(\phi_3-\phi_2)(at+b)^2-b(2k+4-\phi_1)}{(k+2)^2}\right]\frac{1}{t^2},\nonumber\\
\textbf{SEC} &:& \rho+3p =\frac{6}{1-4\alpha^2}\left[\frac{3(\phi_3-3\phi_2)(at+b)^2-b(2k+4-3\phi_1)}{(k+2)^2}\right]\frac{1}{t^2},\nonumber\\
&&~~~~~~~~~ - \frac{4}{1-2\alpha}e^{-\frac{6akt}{k+2}}t^{-\frac{6bk}{k+2}}. \nonumber\\
\textbf{DEC} &:& \rho-p= \frac{6}{1-4\alpha^2}\left[\frac{3(\phi_3+\phi_2)(at+b)^2-b(2k+4+\phi_1)}{(k+2)^2}\right]\frac{1}{t^2},\nonumber\\
&&~~~~~~~~~ + \frac{4}{1-2\alpha}e^{-\frac{6akt}{k+2}}t^{-\frac{6bk}{k+2}}. \nonumber
\end{eqnarray}

The curves of the energy conditions will remain intermediate between the two extremes cases: $a=0$ case and $b=0$ case. It is to mention here that the energy conditions of the two extreme cases can well be recovered by using these extreme values of the HSF parameters.

\section{Summary and Conclusion}
In this work, we have developed a general formalism to investigate Bianchi $VI_h$ universe in an extended gravity theory where the geometrical part of the action integral is modified. The Ricci Scalar $R$ is replaced by a rescaled functional $\lambda(R+T)$ assuming the geometry to  couple with a bit of matter minimally. The motivation behind this is to obtain a set of field equations that can look like the Einstein Field equations(EFE) with a time varying cosmological constant. However, it is not possible to reduce the field equations to EFE because of the non vanishing nature of the scaling constant $\lambda$. Keeping in view of the recent observations predicting an accelerated universe at a late epoch that signals a possible transition from an initial state of deceleration, we employ a hybrid scale factor in the present work. The HSF simulates a signature flipping behaviour of DP. The parameters of the HSF as have been constrained from some recent estimates of transition redshift. The HSF contains two factors, an exponential and a power law functions of the scale factor. While the power law factor dominates at an early cosmic epoch, the exponential part dominates at the late times.

Within the formalism developed here, we have derived the general expressions of the dynamical features of an anisotropic universe using the HSF. These expressions are more general in the sense that, the power law and exponential behaviour appear as the two extreme cases. The dynamical behaviour of the properties remain intermediate to these two extreme cases. In order to assess the dynamical aspects of the model, we have plotted the EoS parameter $\omega$  as a function of redshift and compared its behaviour with that of some well known EoS parametrizations. The EoS evolves with redshit and behaves as a pure cosmological constant at late phase of cosmic time. The rate of dynamical evolution is greatly affected with the change in the HSF parameters. However in the present work with parameters constrained from transition redshift, we obtain a quintessence like behaviour. This shows that the modified gravity theory reproduces quintessence phase of evolution. \\

We have also calculated the energy conditions for the present model that can be suitably reduced to the results already obtained in some earlier works. Since the present approach with a hybrid scale factor appears to be more general than that with a power law or an exponential scale factor, the results of the present study may be more natural and be closer to observations.

\end{document}